\documentclass[10pt, conference]{IEEEtran}
\IEEEoverridecommandlockouts
\usepackage{cite}
\usepackage{amsmath,amssymb,amsfonts}
\usepackage{booktabs,array,makecell,graphicx}
\usepackage{algorithmic}
\usepackage{graphicx}
\usepackage{textcomp}
\usepackage{xcolor}
\usepackage{booktabs}
\usepackage{supertabular}
\usepackage{dirtytalk}
\usepackage[colorlinks,linkcolor=red,anchorcolor=green,citecolor=blue]{hyperref}
\usepackage{threeparttable}
\usepackage{multicol}
\usepackage{multirow}
\usepackage{tikz}

\def\BibTeX{{\rm B\kern-.05em{\sc i\kern-.025em b}\kern-.08em
    T\kern-.1667em\lower.7ex\hbox{E}\kern-.125emX}}

\begin{document}

\newcommand{\TODO}[1]{\textcolor{red}{#1}}
\newcommand{\stef}[1]{\textcolor{black}{#1}}
\newcommand{\stefnew}[1]{\textcolor{black}{#1}}
\newcommand{\chao}[1]{\textcolor{black}{#1}}
\newcommand{\mv}[1]{\textcolor{black}{#1}}
\newcommand{\xyi}[1]{\textcolor{black}{#1}}
\newcommand{\rg}[1]{\textcolor{black}{#1}}
\newcommand{\nsm}[1]{\textcolor{black}{#1}}
\newcommand{\stefnewnew}[1]{\textcolor{black}{#1}}
\newcommand{\stefFinal}[1]{\textcolor{black}{#1}}
\newcommand{\steflast}[1]{\textcolor{black}{#1}}
\newcommand{\steflastest}[1]{\textcolor{black}{#1}}
\newcommand{\stefPostSubm}[1]{\textcolor{black}{#1}}

\newcommand{\stefRVSD}[1]{\textcolor{black}{#1}}
\newcommand{\chaoRVSD}[1]{\textcolor{black}{#1}}

\newcommand{\WIP}[1]{\textcolor{gray}{#1}}

\title{\huge Precision-Scalable Microscaling Datapaths with \\Optimized Reduction Tree for Efficient NPU Integration}

\author{
    Stef Cuyckens*, Xiaoling Yi*, Robin Geens, Joren Dumoulin, Martin Wiesner, Chao Fang\textsuperscript{\textdagger}, Marian Verhelst \\
	\IEEEauthorblockA{
		ESAT-MICAS, KU Leuven, Belgium
    }
    \IEEEauthorblockA{
		Email: \{stef.cuyckens, xiaoling.yi\}@esat.kuleuven.be
    }
}
\maketitle
\vspace{-0.4cm}

\maketitle
\renewcommand{\thefootnote}{}
\footnotetext{$^*$Both authors contributed equally to this work. $^{\dag}$Corresponding author. This project has been partly funded by the European Research Council (ERC) under grant agreement No. 101088865, the European Union’s Horizon 2020 program under grant agreement No. 101070374, the Flanders AI Research Program, Research Foundation Flanders (FWO) under grant No. 1S37125N, and KU Leuven.}


\begin{abstract}

\chao{Emerging continual learning applications necessitate next-generation neural processing unit (NPU) platforms to support both training and inference operations.}
\chao{The promising Microscaling (MX) standard enables narrow bit-widths for inference and large dynamic ranges for training. However, existing MX multiply-accumulate (MAC) designs face a critical trade-off: integer accumulation requires expensive conversions from narrow floating-point products, while FP32 accumulation suffers from quantization losses and costly normalization.}
\chao{To address these limitations, we propose a hybrid precision-scalable reduction tree for MX MACs that combines the benefits of both approaches, enabling efficient mixed-precision accumulation with controlled accuracy relaxation.}
Moreover, we integrate an 8x8 array of these MACs into the \chao{state-of-the-art} \stefPostSubm{(SotA)}
NPU integration platform, SNAX, to provide efficient control and data transfer to our optimized precision-scalable MX datapath. 
We evaluate our design both on MAC and system level and compare it to the SotA. \stefPostSubm{Our integrated system achieves an energy efficiency of 657, 1438-1675, and 4065 GOPS/W, respectively, for MXINT8, MXFP8/6, and MXFP4, with a throughput of 64, 256, and 512 GOPS.}

\end{abstract}

\vspace{-0.2cm}
\section{Introduction}

The rising demand for edge applications capable of continual learning, including robotics, wearable health monitors, and autonomous vehicles \cite{tahir2025edge,sharma2022enabling,CL_driving,pique2022controlling}, requires systems that can adapt to changing environments while staying energy- and area-efficient and meeting tight latency constraints \cite{zhu2024device,ogbogu2023energy,Ekya_gpu,gpu2}. To reduce system-on-chip (SoC) area and cost, both training and inference workloads should be supported by a common compute fabric \cite{training_inference_together,training_supercomputer,liu2024inspire,bai2025npu,taxoNN}. In this unified training–inference NPU platform, the challenge lies in realizing one multiply-accumulate (MAC) array capable of executing both tasks with differing precision requirements, while providing streamlined control logic and high-bandwidth data streaming to sustain peak throughput \cite{Huang2024,PS-MX_MAC}.

Inference workloads typically rely on compact integer formats such as INT8 or INT4 to minimize hardware cost and energy \cite{9610618, liu2024spark, chen2023m4bram}. Training workloads, in contrast, require a much larger dynamic range to maintain model convergence \cite{lu2020evaluations}, which is traditionally provided by FP32 arithmetic. Using high-precision floating-point for all operations is prohibitively expensive in area and power, while using low-precision integer formats for training can lead to significant accuracy loss.

\begin{figure}
    \centering
    \includegraphics[width=0.9\linewidth]{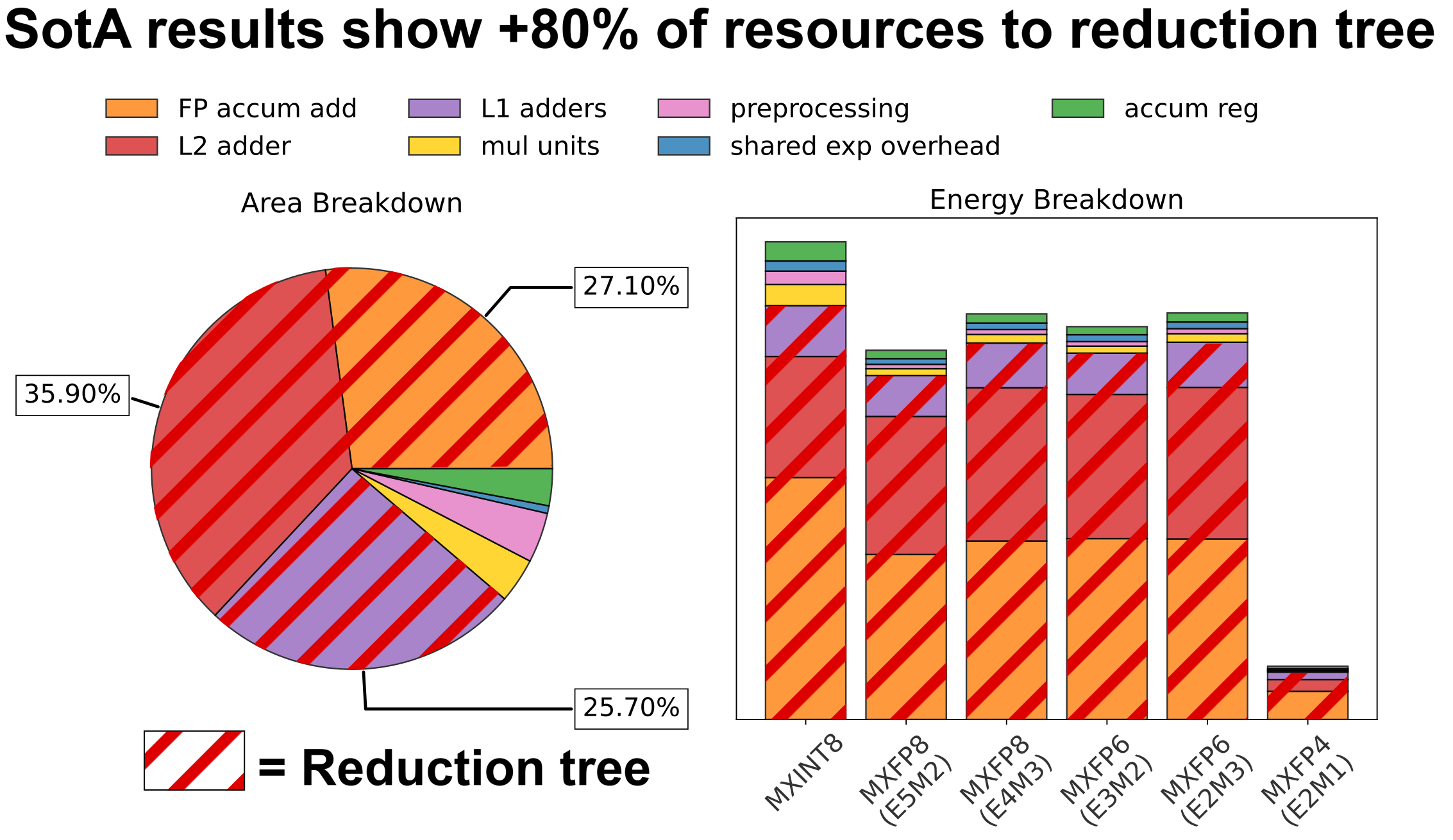}
    \vspace{-0.3cm}
    \caption{Resource breakdown of the state-of-the-art precision-scalable Microscaling (MX) multiply-accumulate (MAC) unit \cite{PS-MX_MAC}, \steflast{where more than $80\%$ of the resources go to the reduction tree.}}
    \label{fig_intro_breakdown}
    \vspace{-0.6cm}
\end{figure}

Microscaling (MX) data types bridge this gap by grouping elements under a shared exponent, thereby preserving dynamic range with narrower element formats such as FP8, FP6, FP4, or INT8 \cite{MX_standard,rouhani2023ocp}. The MX data types thus enable a precision-scalable MAC array that can select data types dynamically, allowing higher precision during training, lower precision during inference, and reduced precision for tasks that tolerate accuracy degradation \cite{MX_standard, PS-MX_MAC}. These capabilities improve energy efficiency and adaptability, making MX well-suited for edge continual learning \cite{tseng2025trainingllmsmxfp4,chen2025oscillationreducedmxfp4trainingvision}.

Despite these advantages, implementing MX-compatible MAC units introduces significant design challenges. Supporting multiple MX data types makes the MAC compute elements resource-heavy, with prior work showing that the accumulation tree dominates both the area ($88.7\%$) and energy ($\approx85\%$) of MX MACs \cite{PS-MX_MAC}, shown in Fig.~\ref{fig_intro_breakdown}. The overhead arises from the need to align products with shared exponents before accumulation: converting the narrow floating-point products into wide integers incurs costly format conversions \cite{MXDotP}, while FP32 accumulation requires expensive normalization logic and incurs quantization losses \cite{PS-MX_MAC}.

While precision-scalable MX MAC-level optimization is critical, the energy efficiency of an NPU is ultimately limited by system-level factors, such as the memory access bottleneck \cite{yi2025opengemm, fang2025anda, yi2022nnasim}. State-of-the-art (SotA) integration framework SNAX \cite{antonio2025open} addresses this issue by providing efficient data management through dedicated data streamers \cite{yi2025datamaestro}. However, these data streamers are provisioned for static worst-case bandwidth, which leads to partially utilized memory channels that dissipate dynamic power and aggravate bank contention during lower-precision operations, a limitation that becomes especially pronounced \steflast{when utilizing the SotA precision-scalable MX MAC units from \cite{PS-MX_MAC}.}

\steflast{To overcome these limitations, this work proposes precision-scalable MX datapaths with an optimized reduction tree and efficient NPU integration with the following key contributions:
\begin{itemize}
    \item At the arithmetic unit level, this work proposes a hybrid precision-scalable reduction tree for MX MACs that combines the key advantages of existing approaches: it leverages the ability of integer accumulation to skip costly normalization, while exploiting the reduced adder width of floating-point accumulation to lower hardware cost.
    \item This implementation is optimized further by relaxing the accumulation accuracy, lowering hardware costs further, while keeping the accuracy loss in check.
    \item At the NPU level, an 8×8 array of the proposed MACs is integrated into the SNAX NPU platform and its data-streaming infrastructure is adapted to efficiently handle the dynamic bandwidth requirements of different MX data types.
\end{itemize}}

\steflast{We evaluate the design at both the MAC and system levels and compare it against SotA MX MAC implementations, ultimately outperforming the previous SotA \cite{PS-MX_MAC} in energy efficiency by a factor of $1.59\times$, $3.05\times$-$3.21\times$, and $1.13\times$, respectively, for MXINT8, MXFP8/6, and MXFP4.}


\begin{figure}
    \centering
    \includegraphics[width=0.78\linewidth]{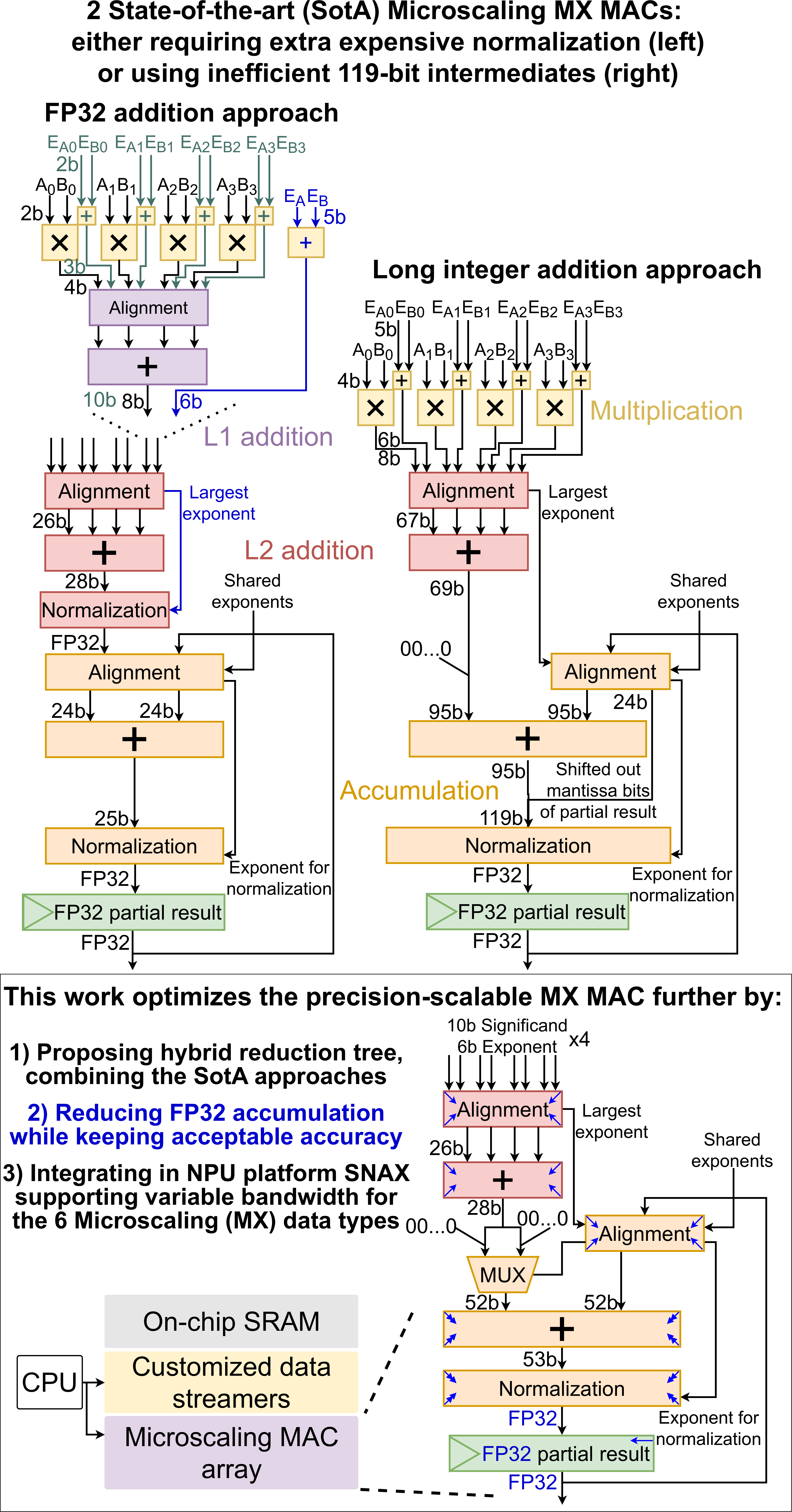}
    \caption{Overview and issues of the state-of-the-art reduction trees for MX MAC implementations\steflast{: FP32 addition\cite{PS-MX_MAC}, and Long integer addition\cite{MXDotP}.} Followed by the solutions proposed in this work. 
    }
    \label{fig_intro}
    \vspace{-0.60cm}
\end{figure}

\section{Background and Motivation} \label{Sec2}
\subsection{MX Format and SotA MX Datapath} \label{Sec2a}
The MX standard \cite{MX_standard} defines six data types, shown in Table~\ref{table_MX}:
MXINT8, MXFP8 E5M2, MXFP8 E4M3, MXFP6 E3M2, MXFP6 E2M3, and MXFP4 E2M1, where the notation $ExMy$ denotes $x$ exponent bits and $y$ mantissa bits. Data is organized into vectors of \mv{(multiples of)} 32 elements, each sharing an 8-bit exponent. The 2-level exponent scheme, consisting of the shared exponent and individual exponents of the MXFP formats, provides a balance between hardware efficiency and dynamic range that is needed to effectively train NN with low-bitwidth quantizations at the edge \cite{MX_standard, PS-MX_MAC}.

The SotA precision-scalable MX MAC from \cite{PS-MX_MAC} is the first to dynamically support all six MX data types within a single INT–FP mixed datapath. This is achieved by splitting the MAC into 2-bit multiply units to handle the 2-bit significands of MXFP4. The MAC has three modes: MXINT8, MXFP8/6, and MXFP4, which sum 1, 4, or 8 products, respectively, before accumulating with a stored FP32 partial result. Fig.~\ref{fig_intro} (left) provides an overview.
Exponent handling occurs at three levels of the adder tree. In MXFP4 (E2M1) mode, each product carries a 3-bit exponent (sum of two 2-bit input exponents), and the resulting 4-bit products (1.M or 0.M) are aligned by exponent prior to level-1 (L1) addition (purple). The L1 output is a 10-bit significand in MXFP4 mode, whereas MXFP8/6 and MXINT8 modes produce 8-bit results, as in conventional precision-scalable MACs \cite{ST}.
At level-2 (L2, red), MXFP8/6 mode also requires exponent alignment: four 8-bit significands, each with up to a 6-bit exponent (from adding MXFP8 E5M2 exponents), are right-shifted relative to the largest exponent. To preserve accuracy, significands are extended to 26 bits, preventing information loss unless alignment shifts exceed this width. The 26-bit width matches FP32’s 24-bit precision, with two guard bits to handle subnormals. The result is then normalized to FP32 and accumulated with the stored FP32 partial sum in level-3 (orange), where shared exponents are applied. After accumulating across cycles, the final FP32 result is quantized back to the target MX format.
Because \cite{PS-MX_MAC} uses 64-element, 8×8 groups with a shared 8-bit exponent, quantization requires all 64 unquantized MAC outputs to compute the correct shared exponent before quantizing to an MX type. These quantized results feed into the next neural network layer. We refer to the adder tree in \cite{PS-MX_MAC} as the FP32-addition approach.

Another SotA work \cite{MXDotP} proposes a different adder tree design (Fig.~\ref{fig_intro}, right). It presents a precision-scalable MX MAC supporting only MXFP8 E5M2 and MXFP8 E4M3. Since MXFP4 is not supported, their adder tree starts at the L2 adder level from \cite{PS-MX_MAC}. Inputs use 8-bit significands with up to 6-bit exponents. Instead of aligning to the highest exponent, they align to a common anchor, place each significand in a 67-bit integer, and add them without bit loss.
For accumulation with an FP32 partial result, they employ early-accumulation \cite{early-accum}, which aligns the FP32 result to the shared MXFP8 exponents, removing the need for normalization and \stefPostSubm{alignment after the addition of the products}. The FP32 mantissa is aligned to the 95-bit product-sum\stefPostSubm{, which was extended with 26 zero-bits,} for lossless accumulation when the FP32 exponent is larger. If the FP32 exponent is much smaller, shifted-out mantissa bits are reattached with proper sign handling, ensuring correctness even for very small sums.
This approach enables accurate accumulation without normalizing or aligning the product-sum, but requires a larger accumulation adder and normalization over a wider bitwidth. More details are provided in \cite{MXDotP,early-accum,Huang2024}. We refer to this design as the long integer addition approach.


\begin{table}[]
\centering
\caption{Concrete MX formats specified by \cite{MX_standard}}
\label{table_MX}
\resizebox{0.45\textwidth}{!}{%
\begin{tabular}{@{}c|cccccc@{}} 
\hline
\textbf{MX name} & MXINT8 & \multicolumn{2}{c}{MXFP8} & \multicolumn{2}{c}{MXFP6} & MXFP4 \\ \hline
\textbf{\begin{tabular}[c]{@{}c@{}}Element \\ format\end{tabular}} & INT8 & E5M2 & E4M3 & E3M2 & E2M3 & E2M1 \\ \hline
\textbf{\begin{tabular}[c]{@{}c@{}}No. bits \\ element\end{tabular}} & 8 & 8 & 8 & 6 & 6 & 4 \\ \hline
\multicolumn{1}{@{}c|}{ } & \multicolumn{6}{c@{}}{Block size of 32 elements. Shared exponent of 8 bits.} \\ \hline
\end{tabular}%
}
\vspace{-0.60cm}
\end{table}

\subsection{SNAX Cluster} \label{sec2b}


SNAX \cite{antonio2025open} is a RISC-V compute cluster template for rapid NPU integration. It features a RISC-V RV32IMAFD Snitch \cite{zaruba2020snitch} core that controls the NPU through standard Configuration and Status Register (CSR) instructions. A 128 KiB 32-banked shared scratchpad memory (SPM) with a fully connected crossbar is leveraged to store the operands and provide high bandwidth data access. Dedicated data streamers are designed to provide autonomous and continuous data streams towards the NPU, maximizing the NPU's utilization. A DMA core handles memory transfers between the SPM and external memory, providing 512-bit peak data bandwidth.

To support agile NPU integration, SNAX employs a hybrid data/control coupling strategy to boost the NPU's ease of programmability and efficiency at the system level. A loosely coupled control interface enables efficient kernel offloading without stalling the RISC-V Snitch core, while the CSR-based configuration model abstracts hardware details for ease of programming. In parallel, tightly coupled data streamers \cite{yi2025datamaestro} provide low-latency access to shared memory, accommodating diverse bandwidth demands at design time and a variety of data access patterns at runtime for diverse NPU workloads.

\section{Reduction Tree Design} \label{Sec3}
\subsection{Hybrid Reduction Tree}\label{Sec3a}
Our work builds on the precision-scalable MX MAC architecture from \cite{PS-MX_MAC}, which supports all MX types and thus improves adaptability for training–inference platforms. Our optimizations focus on the L2 adder and accumulation, as these dominate the MAC’s energy and area costs (Fig.~\ref{fig_intro_breakdown}). So, the inputs to our reduction tree design are four 10-bit significands with 6-bit exponents, as discussed in Sec. \ref{Sec2a}, and two 8-bit shared exponents, one for each factor, as in \cite{PS-MX_MAC, MXDotP}.

\begin{figure}
    \centering
    \includegraphics[width=0.78\linewidth]{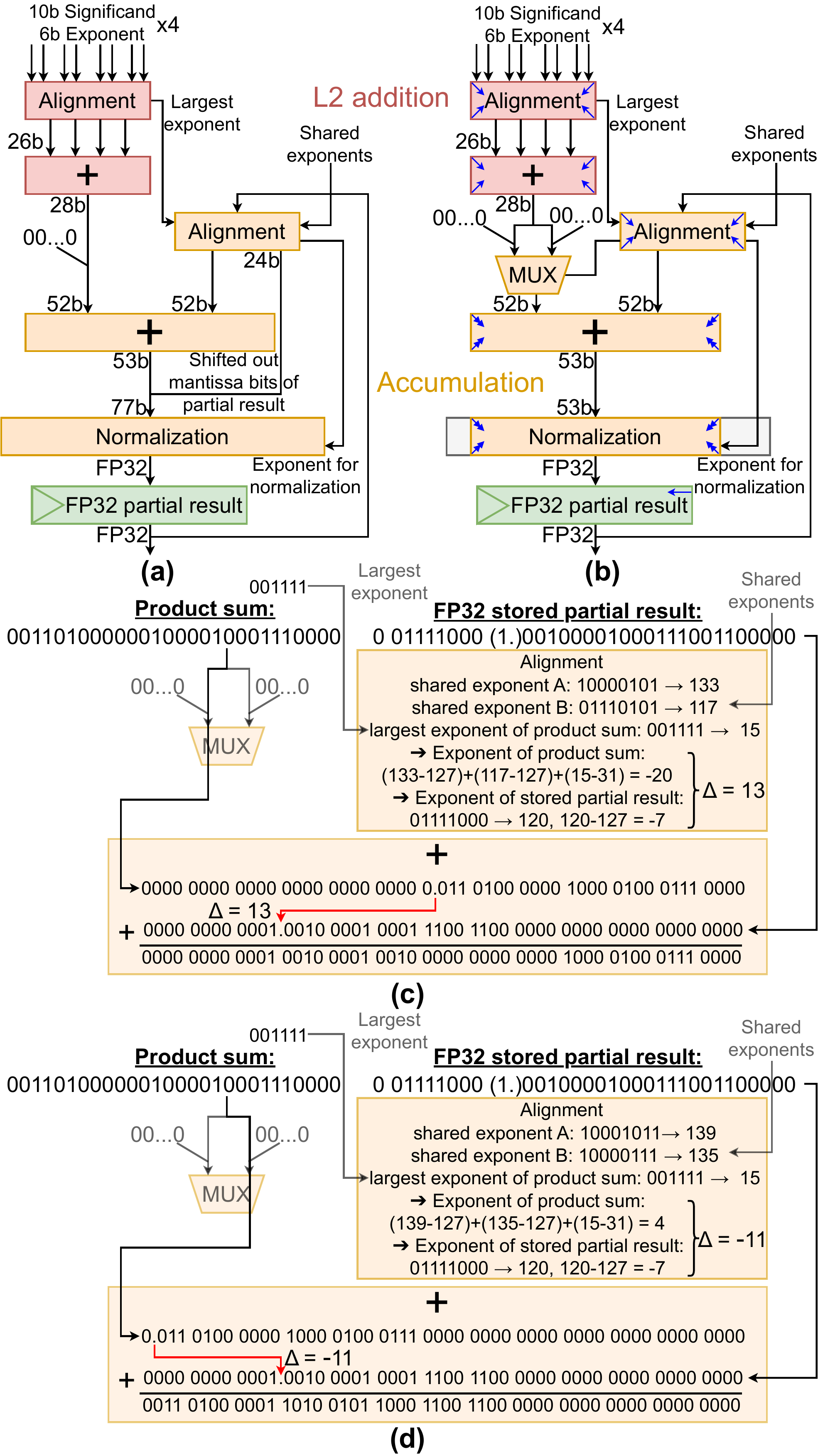}
    \vspace{-0.3cm}
    \caption{First (a) and second (b) iteration of our proposed hybrid reduction tree architecture with examples (c, d) to illustrate the multiplexer in (b).
    }
    \label{hybrid_adder_tree_evolution}
    \vspace{-0.5cm}
\end{figure}

Specifically, we integrate the early-accumulation scheme from \cite{MXDotP, early-accum} 
into the FP32 reduction tree of \cite{PS-MX_MAC}, as shown in Fig.~\ref{hybrid_adder_tree_evolution}a. This enables efficient 28-bit signal widths that avoid adder overprovisioning in the L2 adder while reducing normalization and alignment overhead between the L2 adder and accumulation. The drawback of incorporating early-accumulation is the larger adder and normalization in the accumulation \steflast{compared to the FP32 addition approach \cite{PS-MX_MAC}}.
In Fig.~\ref{hybrid_adder_tree_evolution}a, the 28-bit product-sum is extended by 24 bits to allow the 24-bit significand of the stored FP32 value to be shifted to the left for proper addition. After the addition, the 53-bit output is further extended by 24 bits, as the whole 24-bit significand could be shifted out and needs to be reattached for proper normalization. This results in a normalization that needs to support a 77-bit input.

To address this high normalization overhead, the reduction tree is optimized further, as illustrated in Fig.~\ref{hybrid_adder_tree_evolution}b. Here, we leverage the fact that the 24-bit extension before addition is only needed if the stored FP32 partial result is larger than the product sum and needs to be shifted to the left relative to the product sum. While the 24-bit extension after the addition is only used if the significand of the stored FP32 partial result needs to be shifted to the right. Because only one 24-bit extension can be used at the same time, the normalization can be relaxed to an input width of 53 bits. 
This is implemented by a multiplexer (MUX) that either extends the 28-bit product sum on the left or right, depending on how the significand of the stored FP32 value is shifted in the alignment. The sign handling is not shown for conciseness in Fig.~\ref{hybrid_adder_tree_evolution}. These two accumulation configurations are illustrated with examples in Fig.~\ref{hybrid_adder_tree_evolution}c and \ref{hybrid_adder_tree_evolution}d. 

\subsection{\mv{Accumulation Precision Optimization}}
\label{Sec3b}

When working with floating-point data formats, one of the most important parameters is the accuracy of the result, which depends primarily on the number of mantissa bits used in the computation \cite{Mant_accuracy} and on how many of these are actually kept in the final result. Most works store partial results in FP32, with its 23 mantissa bits of accuracy \cite{PS-MX_MAC, MXDotP}. Efficient implementations avoid excessive computations by keeping the accuracy of the computation in line with the accuracy of the final result. So, \stefPostSubm{as a consequence}, the accuracy of the stored partial results influences the whole adder tree structure. This section proposes to reduce the number of mantissa bits of the stored partial result in our hybrid reduction tree design to further optimize our implementation of the precision-scalable MX MAC.

First, we look at how our reduction tree design is influenced by reducing the number of mantissa bits of the stored partial result. In Fig.~\ref{hybrid_adder_tree_evolution}b, the blue arrows indicate how much the block can scale down; the ones with single-tipped arrows scale down proportionally to the mantissa width, and the bitwidth of the modules with double-tipped arrows decreases by two bits for every removed bit from the mantissa. The L2 alignment and addition scale proportionally because the accuracy of this addition is designed in \cite{PS-MX_MAC} to be the same as the stored partial result. In the accumulation, the alignment of the stored partial result, the addition, and the normalization, all scale twice as much. This is because the bitwidth of the inputs of these modules is found by combining the output width of the L2 addition with the mantissa width of the stored partial result. Previously, 28 bits with the 24-bit significand of the stored partial result to make the 52-bit signal width, where both of these scale with the mantissa width.

\begin{figure}[t]
    \centering
    \includegraphics[width=0.92\linewidth]{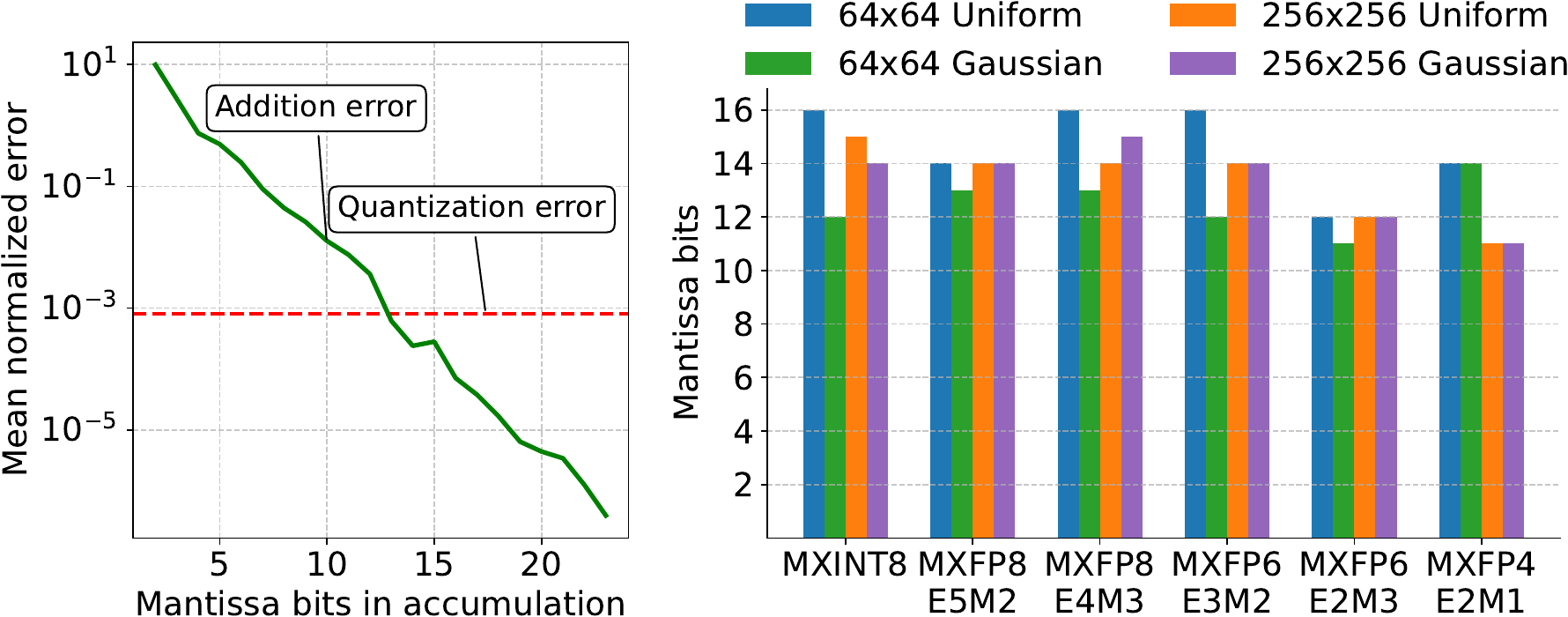}
    \vspace{-0.2cm}
    \caption{Comparing quantization error and addition error for reduced mantissa length in accumulation, both errors are normalized to the result computed in float64, which is also treated as the perfect result when computing the errors: (left) error comparison for MXFP8 E4M3 with matrix sizes of 64x64 and Gaussian distributed matrix elements, (right) the lowest mantissa lengths for which the quantization error is larger than the addition error, for both 64x64 and 256x256 matrix sizes and for uniform and Gaussian distributions of the matrix elements.}
    \label{fig_M_acc}
    \vspace{-0.5cm}
\end{figure}

Second, we check how much the mantissa width of the stored intermediates can be reduced without affecting the accuracy of the final result. As mentioned in Sec. \ref{Sec2a}, the final floating-point result, stored in the accumulation register, is quantized to one of the MX data formats after all the data is accumulated. This quantization was implemented as in \cite{MX_standard}, where the new shared scale is computed by finding the largest element in the MX group and dividing each of them by the shared scale before quantizing the individual elements to INT8, FP8, FP6, or FP4.
This naturally results in a certain quantization error. This quantization error is unavoidable, as otherwise the computation of the next layer would need to be done in FP32, requiring much more resources than the MX data formats. Now, to estimate how much the mantissa width of the stored intermediate results can be reduced without compromising the accuracy of addition, we compare the addition error against this quantization error. Our assumption is that, as long as the quantization error exceeds the addition error, the latter becomes negligible. This condition allows us to reduce the mantissa width until the two error sources become comparable in magnitude. 

Because the target application for our training-inference NPU platform is continual learning, we cannot optimize the mantissa width for a predefined workload. Instead, the aim is to compare the addition and quantization errors for a general workload. We emulate this general workload in two ways: 1) by using a uniform distribution for the elements of the MX groups and the shared exponents, and 2) by using a Gaussian distribution. The shared exponents of the uniform distribution are limited to a practical range of -32 to 32 to avoid overflow, and the standard deviation of the Gaussian distribution is found by setting $6\sigma = 2^{32}$.

In Fig.~\ref{fig_M_acc} (left), the addition and quantization error of matrix multiplication with size 64x64 for MXFP8 E4M3 quantization is shown. On the x-axis, the mantissa width is varied between 2 bits and 23 bits, where both the stored partial results and the computation are done in this mantissa length. The errors are calculated relative to the FP64 result and are normalized by this FP64 value. This normalization prevents the errors from being dominated by a few large values, providing instead a measure of the relative error across all data points. In Fig.~\ref{fig_M_acc}, the quantization error and addition error become proportional for a mantissa width of 13 bits.
Fig.~\ref{fig_M_acc} (right) reports these \stefPostSubm{critical} mantissa widths \stefPostSubm{for which the addition error equals the quantization error} for the different MX data formats, different matrix sizes, and for the two input distributions. The highest mantissa is chosen for the hardware implementation, as this provides accurate addition even in the worst case. To this end, our design uses a mantissa width of 16-bit.

\section{NPU Integration} \label{Sec4}

The precision-scalable MX MACs are organized into an 8×8 spatial array, referred to as the MX tensor core, to enable spatial data reuse and parallel execution of general matrix multiplication (GeMM) workloads. This section first presents the architecture of the MX tensor core (Sec. \ref{mx_tensor_core}) and then describes its integration into the SotA NPU platform SNAX \cite{antonio2025open}, including the flexible control of the MX tensor core (Sec. \ref{snax_control}) and efficient data streaming (Sec. \ref{snax_data}).



\begin{figure}
    \centering
    \includegraphics[width=\linewidth]{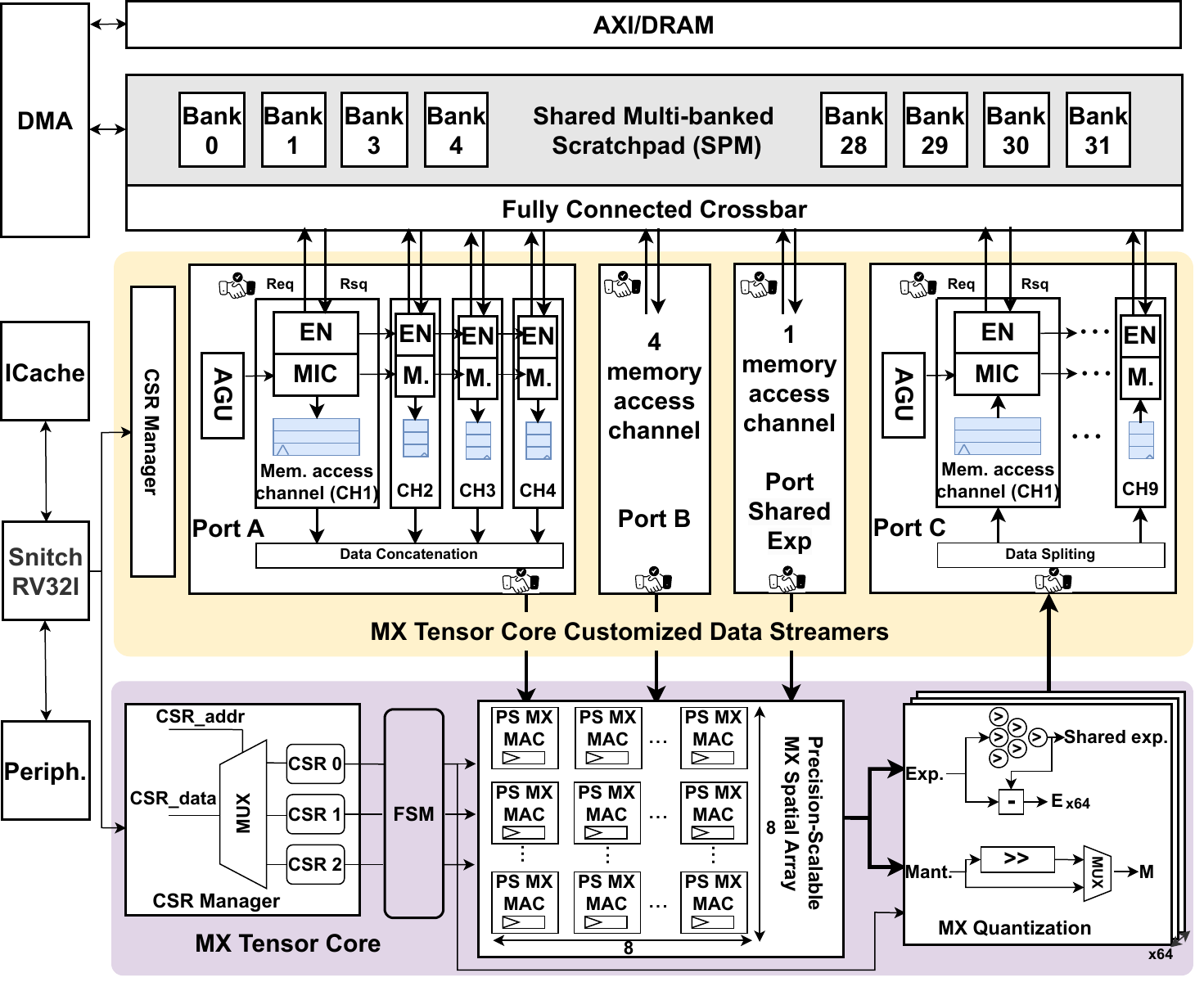}
    \vspace{-0.3cm}
    \caption{System architecture overview with precision-scalable MX tensor core integration. }
    \label{fig:sys_arch}
    \vspace{-0.5cm}
\end{figure}

\subsection{MX Tensor Core}
\label{mx_tensor_core}

As shown in Fig.~\ref{fig:sys_arch} (bottom), the MX tensor core consists of three primary submodules: a flexible finite state machine (FSM), a precision-scalable MX spatial array, and a SIMD-based quantization unit. The FSM provides microarchitectural control and handshake signals. It configures the precision mode for the compute array and orchestrates the entire GeMM execution process by generating control signals based on the matrix size. The spatial array comprises 64 precision-scalable MX MAC units arranged in an 8 × 8 two-dimensional mesh, enabling both horizontal and vertical data reuse \cite{moons201714,ueyoshi2022diana}. Depending on the selected precision mode, the array performs the GeMM operation on two 8 × 8 matrices in 8, 2, or 1 cycles for INT8, FP8/FP6, and FP4 modes, respectively \cite{PS-MX_MAC}. Located downstream of the spatial array, the SIMD quantization unit processes the 64 floating-point outputs and converts them into any supported MX numerical format. This separation of control, computation, and quantization enables the tensor core to dynamically adapt to different precision and matrix size requirements while sustaining high computational throughput.

\subsection{Unified Control Through CSR}
\label{snax_control}

The MX tensor core is programmed by the Snitch RISC-V processor via standard CSR writes, providing a configuration bandwidth of 32 bits per cycle. As illustrated in Fig.~\ref{fig:sys_arch} (bottom left), a dedicated CSR manager exposes three registers that form a uniform programming interface. As shown in Table~\ref{tab:csr}, CSR 0 selects the numerical precision for both the spatial array and the quantization unit, whereas CSR 1 and CSR 2 define the accumulation dimension and the matrix-tile dimension, respectively. By issuing CSR write instructions, the Snitch core dynamically configures the MX tensor core FSM, adapting the compute array to the desired precision, accumulation depth, and matrix size at runtime.

\begin{table}[t]
  \centering
  \caption{CSRs for MX tensor core configuration}
  \label{tab:csr}
  \begin{tabular}{@{}ll@{}}
    \toprule
    Register & Function \\ \midrule
    CSR0 & MX array and quantization precision mode selection \\
    CSR1 & Accumulation times for one result output \\
    CSR2 & Block matrix size: the row/column dimension of the matrix tile \\
    \bottomrule
  \end{tabular}
  \vspace{-0.5cm}
\end{table}

\subsection{Streamlined Data Supply Through Dynamic Streamers}
\label{snax_data}

As shown in Fig.~\ref{fig:sys_arch} (middle), we employ separate data streamers with independent memory access channels that satisfy the bandwidth demands of the MX tensor core. To accommodate the MX tensor core’s varying bandwidth requirements across different precision modes, we extend the SNAX data streamers \cite{yi2025datamaestro} with dynamic channel gating. At run-time, the streamer activates only the subset of memory access channels required by the current precision mode: MXINT8, MXFP8, MXFP6, and MXFP4 modes require 1, 4, 3, and 4 channels, respectively, according to the computation characteristics in different precision modes \cite{PS-MX_MAC}. This selective activation reduces unnecessary memory traffic and mitigates both energy consumption and bank contention in the memory subsystem.
Furthermore, each precision mode stores matrix tiles in a data layout optimized for its operand width and exhibits distinct access patterns from the MX tensor core’s perspective. A programmable address-generation unit (AGU) within the data streamers can be configured by the Snitch core at run-time to support the appropriate data layout and access pattern.
In summary, the enhanced data streamers enable flexible adaptation of bandwidth, data layout, and access patterns to efficiently support multiple MX precision modes.

\section{Experimental Results} \label{Sec5}


We implement the precision-scalable MX MAC, MX tensor core, and enhanced SNAX integration infrastructure at RTL in SystemVerilog. Performance, area, and power are evaluated at MAC and system level using a standard VLSI synthesis flow with Synopsys Design Compiler\textsuperscript{\textregistered} in GlobalFoundries 22FDX\textsuperscript{\textregistered} technology under a nominal supply voltage of $0.8V$ in the typical-typical corner. For accurate power results, we run netlist simulations in Siemens QuestaSim\textsuperscript{\texttrademark} to generate switching activity, and use Synopsys PrimeTime PX\textsuperscript{\texttrademark} for power and energy analysis. Clock gating is applied in the synthesis flow to improve energy efficiency. 

\subsection{MAC Evaluation}

\begin{figure}
    \centering
    \includegraphics[width=0.9\linewidth]{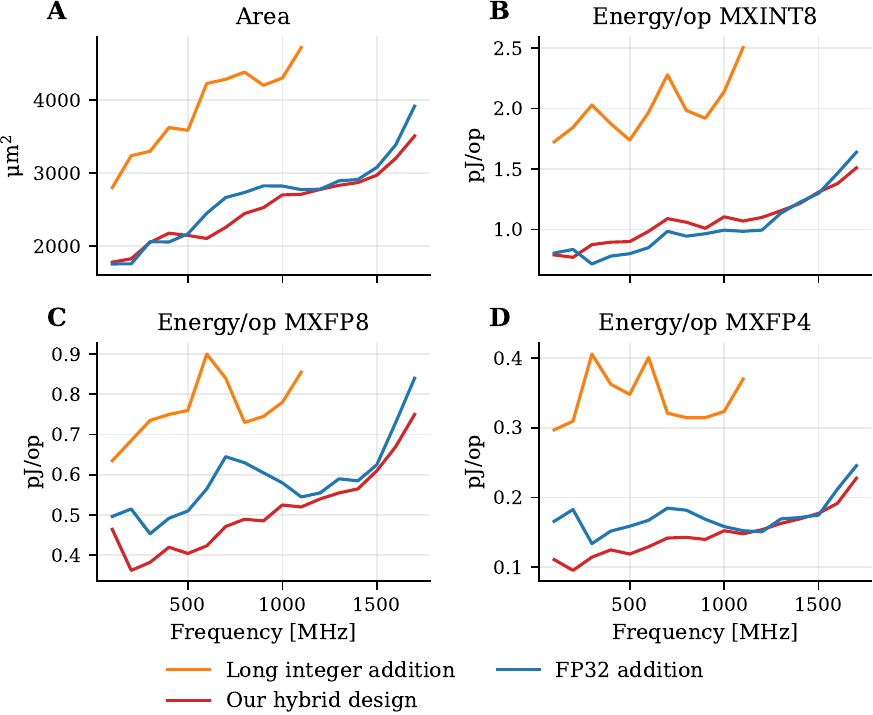}
    \vspace{-0.3cm}
    \caption{MAC-level comparison of our precision-scalable MX MAC design against the SotAs: Long integer addition~\cite{MXDotP} and FP32 addition\cite{PS-MX_MAC}. Each multiplication and addition counts as an operation, resulting in 2 ops/cycle for MXINT8, 8 ops/cycle for MXFP8 and MXFP6, and 16 ops/cycle for MXFP4.}
    \label{fig_MAC_experiment}
    \vspace{-0.2cm}
\end{figure}

This section evaluates our precision-scalable MX MAC with the hybrid reduction tree design and compares it to the two SotA adder tree designs from Sec. \ref{Sec2a}. By substituting the SotA adder tree into our MX MAC \cite{PS-MX_MAC}, we enable a direct comparison of addition strategies. Designs are compared at clock frequencies from 100 MHz to 1800 MHz. While all MX formats are supported, Fig.~\ref{fig_MAC_experiment} shows only MXINT8, MXFP8 E4M3, and MXFP4 for conciseness, as results for other MXFP8 and MXFP6 formats are similar to MXFP8 E4M3. The long integer addition approach \cite{MXDotP} is outperformed by our hybrid design in area and energy across the full frequency range. Our design reaches 1800 MHz, whereas the SotA design reaches only 1100 MHz. Compared to FP32 addition \cite{PS-MX_MAC}, our design is more energy-efficient in MXFP8 and MXFP4 modes below 1 GHz; above this frequency, efficiency is comparable. In MXINT8 mode, the FP32 SotA design remains more energy-efficient. For area efficiency, our design is superior between 500 and 1000 MHz.

\subsection{NPU Integration Evaluation} \label{Sec5c}
We now evaluate the system-level performance of the integrated NPU and also show the area and energy breakdown of the system, operating at a frequency of 500 MHz.

\subsubsection{Performance Evaluation}
\begin{figure}
    \centering
    \includegraphics[width=0.65\linewidth]{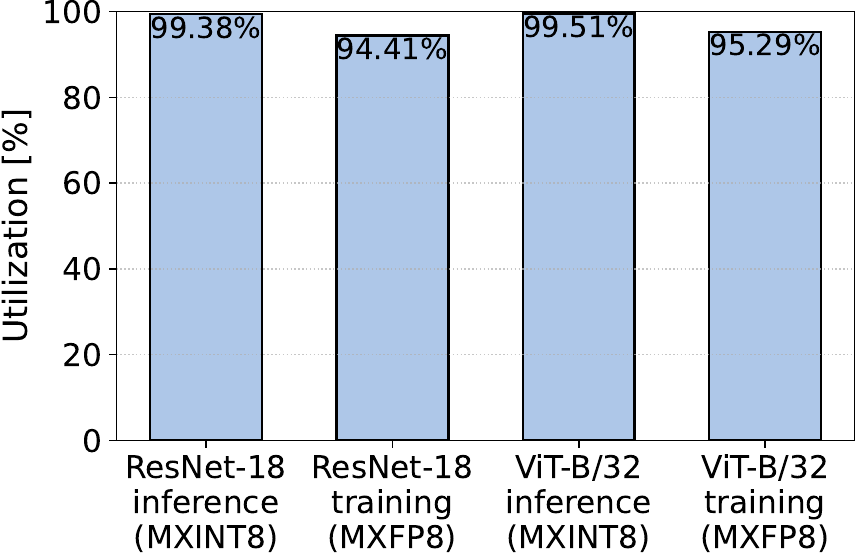}
    \vspace{-0.2cm}
    \caption{Temporal utilization of inference (INT8) and training (FP8 E4M3) workloads on our integrated MX tensor core NPU.}
    \label{fig_util}
    \vspace{-0.4cm}
\end{figure}

We evaluate the MX tensor core on ResNet18 and Vision Transformer for inference and training with batch size 32 \cite{dacapo}. Inference uses INT8, while training uses FP8 E4M3 for its larger dynamic range \cite{noune20228bitnumericalformatsdeep}.
As depicted in Fig.~\ref{fig_util}, the MX tensor core achieves a computational utilization ranging from 94.41\%-99.51\% across these four workloads. This high utilization indicates that the system successfully minimizes control and memory bottlenecks, allowing the core to operate near its theoretical peak throughput.

\subsubsection{Area and Power Analysis}

\begin{figure}
    \centering
    \includegraphics[width=0.85\linewidth]{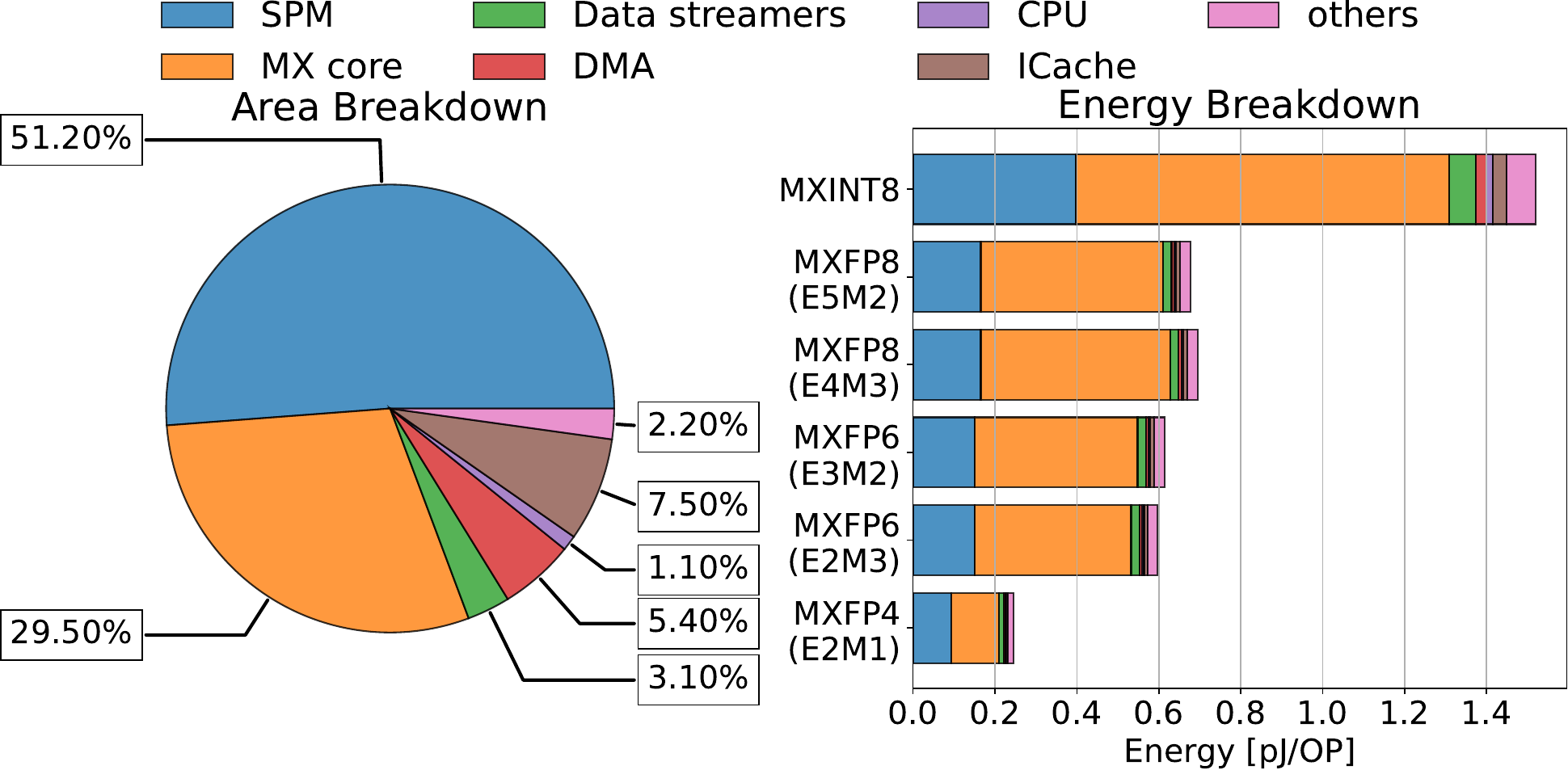}
    \vspace{-0.3cm}
    \caption{Area and power breakdown of our integrated NPU.}
    \label{fig_area_power_breakdown}
    \vspace{-0.4cm}
\end{figure}

\steflast{Fig.~\ref{fig_area_power_breakdown} shows a detailed area and energy breakdown. The system occupies 0.60$mm^2$, mostly taken by the multi-banked SPM and data supply, leaving 29.5\% for the MX tensor core. In contrast, the MX core dominates the energy breakdown due to dynamic channel gating in the data streamers and clock gating in the synthesis flow, which enhance the energy efficiency of sequential logic like the SPM and data supply. Our code is available at \cite{PS_MX_repo}.}



\subsection{SotA Comparison}
\steflast{We compare our work with the SotA in Table~\ref{SotA_table}. Compared to other MX compute cores \cite{MXDotP,PS-MX_MAC}, our system outperforms both in energy efficiency. Here, the improvements compared to \cite{PS-MX_MAC} are larger than expected from Fig.~\ref{fig_MAC_experiment}, \steflastest{due to this work using register optimization setting and more relaxed constraints in the synthesis flow}. Compared to a non-precision-scalable GeMM core for INT8 \cite{yi2025opengemm}, we see that the precision-scalability and MX support of our design do give rise to a serious gap in energy- and area-efficiency.}

\begin{table}[t]
\centering
\caption{State-of-the-art (SotA) comparison.}
\label{SotA_table}

\newcommand{\tablescale}{0.48} 

\resizebox{\tablescale\textwidth}{!}{%
\begin{tabular}{|l|c|c|c|c|}
\hline
\textbf{} 
 & \makecell{\cite{MXDotP}} 
 & \makecell{\cite{PS-MX_MAC}} 
 & \makecell{\cite{yi2025opengemm}} 
 & \makecell{Our Work} \\
\hline
\textbf{\makecell{Frequency (MHz)}} & 1000 & 400 & 200 & 500\\
\hline
\textbf{\makecell{Technology (nm)}} & 12 & 16 & 16 & 22\\
\hline
\textbf{\makecell{Area (mm$^2$)}} & 0.59 & 8.92 & 0.62 & 0.60\\
\hline
\textbf{\makecell{Area/MAC (um$^2$)}} & 3150 & 2080 & 144 & 2766\\
\hline
\textbf{\makecell{Supported\\precisions}} & MXFP8 & \makecell{MX (INT8, FP8,\\ FP6, FP4)} & INT8 & \makecell{MX (INT8, FP8,\\ FP6, FP4)}\\
\hline
\textbf{\makecell{Throughput\\ (GOPS)}} & 102 & / & 204 & \makecell{MXINT8: 64,\\MXFP8/6: 256,\\MXFP4: 512}\\
\hline
\textbf{\makecell{Energy\\efficiency\\(GOPS/W)}} & 356 & \makecell{MXINT8: 412,\\ MXFP8/6: 472–521,\\ MXFP4: 3597} & 4680 & \makecell{MXINT8: 657,\\MXFP8/6: 1438–1675,\\MXFP4: 4065}\\
\hline
\textbf{\makecell{NPU integrated?}} & $\checkmark$ & $\times$ & $\checkmark$ & $\checkmark$ \\
\hline
\textbf{\makecell{Precision-scalable?}} & $\times$ & $\checkmark$ & $\times$ & $\checkmark$ \\
\hline
\end{tabular}%
}
\vspace{-0.5cm}
\end{table}

\section{Conclusion}
In this work, we address the challenges of developing an efficient system implementation for a precision-scalable Microscaling compute core: (1) by developing a new reduction tree approach that takes inspiration from the SotA implementations, (2) further optimizing the reduction tree by carefully reducing the accumulation precision, and (3) integrating our compute units in the NPU platform, SNAX, and extending the SNAX implementation with dynamic bandwidth depending on the precision of the computation. 
\steflast{Our integrated system outperforms the previous SotA \cite{PS-MX_MAC} in energy efficiency by a factor of $1.59\times$, $3.05\times$-$3.21\times$, and $1.13\times$, respectively, for MXINT8, MXFP8/6, and MXFP4. While also being efficiently integrated in the NPU platform, SNAX.}

\clearpage
\bibliographystyle{IEEEtran}
\bibliography{refs}

\begin{thebibliography}{10}
\providecommand{\url}[1]{#1}
\csname url@samestyle\endcsname
\providecommand{\newblock}{\relax}
\providecommand{\bibinfo}[2]{#2}
\providecommand{\BIBentrySTDinterwordspacing}{\spaceskip=0pt\relax}
\providecommand{\BIBentryALTinterwordstretchfactor}{4}
\providecommand{\BIBentryALTinterwordspacing}{\spaceskip=\fontdimen2\font plus
\BIBentryALTinterwordstretchfactor\fontdimen3\font minus \fontdimen4\font\relax}
\providecommand{\BIBforeignlanguage}[2]{{%
\expandafter\ifx\csname l@#1\endcsname\relax
\typeout{** WARNING: IEEEtran.bst: No hyphenation pattern has been}%
\typeout{** loaded for the language `#1'. Using the pattern for}%
\typeout{** the default language instead.}%
\else
\language=\csname l@#1\endcsname
\fi
#2}}
\providecommand{\BIBdecl}{\relax}
\BIBdecl

\bibitem{tahir2025edge}
N.~Tahir \emph{et~al.}, ``Edge computing and its application in robotics: A survey,'' \emph{Journal of Sensor and Actuator Networks (JSAN)}, vol.~14, no.~4, p.~65, 2025.

\bibitem{sharma2022enabling}
D.~Sharma \emph{et~al.}, ``Enabling inference and training of deep learning models for ai applications on iot edge devices,'' in \emph{Artificial Intelligence-based Internet of Things Systems}, 2022, pp. 267--283.

\bibitem{CL_driving}
H.~Yang \emph{et~al.}, ``Human-guided continual learning for personalized decision-making of autonomous driving,'' \emph{IEEE Transactions on Intelligent Transportation Systems (TITS)}, vol.~26, no.~4, pp. 5435--5447, 2025.

\bibitem{pique2022controlling}
F.~Piqu{\'e} \emph{et~al.}, ``Controlling soft robotic arms using continual learning,'' \emph{IEEE Robotics and Automation Letters (RAL)}, vol.~7, no.~2, pp. 5469--5476, 2022.

\bibitem{zhu2024device}
S.~Zhu \emph{et~al.}, ``On-device training: A first overview on existing systems,'' \emph{ACM Transactions on Sensor Networks (TOSN)}, vol.~20, no.~6, pp. 1--39, 2024.

\bibitem{ogbogu2023energy}
C.~Ogbogu \emph{et~al.}, ``Energy-efficient machine learning acceleration: from technologies to circuits and systems,'' in \emph{2023 IEEE/ACM International Symposium on Low Power Electronics and Design (ISLPED)}.\hskip 1em plus 0.5em minus 0.4em\relax IEEE, 2023, pp. 1--8.

\bibitem{Ekya_gpu}
R.~Bhardwaj \emph{et~al.}, ``Ekya: Continuous learning of video analytics models on edge compute servers,'' in \emph{19th USENIX Symposium on Networked Systems Design and Implementation (NSDI)}.\hskip 1em plus 0.5em minus 0.4em\relax Renton, WA: USENIX Association, Apr. 2022, pp. 119--135.

\bibitem{gpu2}
Y.~Kong \emph{et~al.}, ``Edge-assisted on-device model update for video analytics in adverse environments,'' in \emph{Proceedings of the 31st ACM International Conference on Multimedia (MM)}, 2023, pp. 9051--9060.

\bibitem{training_inference_together}
S.~Shukla \emph{et~al.}, ``A scalable multi-teraops core for ai training and inference,'' \emph{IEEE Solid-State Circuits Letters (SSCL)}, vol.~1, no.~12, pp. 217--220, 2018.

\bibitem{training_supercomputer}
N.~P. Jouppi \emph{et~al.}, ``A domain-specific supercomputer for training deep neural networks,'' \emph{Communications of the ACM (CACM)}, vol.~63, no.~7, pp. 67--78, 2020.

\bibitem{liu2024inspire}
F.~Liu \emph{et~al.}, ``Inspire: Accelerating deep neural networks via hardware-friendly index-pair encoding,'' in \emph{Proceedings of the 61st ACM/IEEE Design Automation Conference (DAC)}, 2024, pp. 1--6.

\bibitem{bai2025npu}
Y.~Bai \emph{et~al.}, ``Be-npu: A bandwidth-efficient neural processing unit with adaptive processing schemes for reduced off-chip bandwidth demand,'' \emph{IEEE Transactions on Computers (TC)}, 2025.

\bibitem{taxoNN}
R.~Hojabr \emph{et~al.}, ``Taxonn: A light-weight accelerator for deep neural network training,'' in \emph{2020 IEEE International Symposium on Circuits and Systems (ISCAS)}, 2020, pp. 1--5.

\bibitem{Huang2024}
L.~Huang \emph{et~al.}, ``A precision-scalable risc-v dnn processor with on-device learning capability at the extreme edge,'' in \emph{2024 29th Asia and South Pacific Design Automation Conference (ASP-DAC)}, 2024, pp. 927--932.

\bibitem{PS-MX_MAC}
S.~Cuyckens \emph{et~al.}, ``Efficient precision-scalable hardware for microscaling (mx) processing in robotics learning,'' in \emph{IEEE/ACM International Symposium on Low Power Electronics and Design (ISLPED)}, 2025, pp. 1--7.

\bibitem{9610618}
S.~K. Lee \emph{et~al.}, ``A 7-nm four-core mixed-precision ai chip with 26.2-tflops hybrid-fp8 training, 104.9-tops int4 inference, and workload-aware throttling,'' \emph{IEEE Journal of Solid-State Circuits (JSSC)}, vol.~57, no.~1, pp. 182--197, 2022.

\bibitem{liu2024spark}
F.~Liu \emph{et~al.}, ``Spark: Scalable and precision-aware acceleration of neural networks via efficient encoding,'' in \emph{2024 IEEE International Symposium on High-Performance Computer Architecture (HPCA)}.\hskip 1em plus 0.5em minus 0.4em\relax IEEE, 2024, pp. 1029--1042.

\bibitem{chen2023m4bram}
Y.~Chen \emph{et~al.}, ``M4bram: Mixed-precision matrix-matrix multiplication in fpga block rams,'' in \emph{2023 International Conference on Field Programmable Technology (ICFPT)}.\hskip 1em plus 0.5em minus 0.4em\relax IEEE, 2023, pp. 69--78.

\bibitem{lu2020evaluations}
J.~Lu \emph{et~al.}, ``Evaluations on deep neural networks training using posit number system,'' \emph{IEEE Transactions on Computers (TC)}, vol.~70, no.~2, pp. 174--187, 2020.

\bibitem{MX_standard}
\BIBentryALTinterwordspacing
B.~D. Rouhani \emph{et~al.}, ``Microscaling data formats for deep learning,'' 2023. [Online]. Available: \url{https://arxiv.org/abs/2310.10537}
\BIBentrySTDinterwordspacing

\bibitem{rouhani2023ocp}
\BIBentryALTinterwordspacing
B.~D. Rouhani \emph{et~al.}, ``Ocp microscaling formats (mx) specification,'' Open Compute Project, 2023, accessed: March 12, 2025. [Online]. Available: \url{https://www.opencompute.org/documents/ocp-microscaling-formats-mx-v1-0-spec-final-pdf}
\BIBentrySTDinterwordspacing

\bibitem{tseng2025trainingllmsmxfp4}
\BIBentryALTinterwordspacing
A.~Tseng \emph{et~al.}, ``Training llms with mxfp4,'' 2025. [Online]. Available: \url{https://arxiv.org/abs/2502.20586}
\BIBentrySTDinterwordspacing

\bibitem{chen2025oscillationreducedmxfp4trainingvision}
\BIBentryALTinterwordspacing
Y.~Chen \emph{et~al.}, ``Oscillation-reduced mxfp4 training for vision transformers,'' 2025. [Online]. Available: \url{https://arxiv.org/abs/2502.20853}
\BIBentrySTDinterwordspacing

\bibitem{MXDotP}
G.~İslamoğlu \emph{et~al.}, ``Mxdotp: A risc-v isa extension for enabling microscaling (mx) floating-point dot products,'' in \emph{IEEE 36th International Conference on Application-specific Systems, Architectures and Processors (ASAP)}, 2025, pp. 81--84.

\bibitem{yi2025opengemm}
X.~Yi \emph{et~al.}, ``Opengemm: A highly-efficient gemm accelerator generator with lightweight risc-v control and tight memory coupling,'' in \emph{Proceedings of the 30th Asia and South Pacific Design Automation Conference (ASP-DAC)}, 2025, pp. 1055--1061.

\bibitem{fang2025anda}
C.~Fang \emph{et~al.}, ``Anda: Unlocking efficient llm inference with a variable-length grouped activation data format,'' in \emph{2025 IEEE International Symposium on High Performance Computer Architecture (HPCA)}.\hskip 1em plus 0.5em minus 0.4em\relax IEEE, 2025, pp. 1467--1481.

\bibitem{yi2022nnasim}
X.~Yi \emph{et~al.}, ``Nnasim: An efficient event-driven simulator for dnn accelerators with accurate timing and area models,'' in \emph{2022 IEEE International Symposium on Circuits and Systems (ISCAS)}.\hskip 1em plus 0.5em minus 0.4em\relax IEEE, 2022, pp. 2806--2810.

\bibitem{antonio2025open}
R.~A. Antonio \emph{et~al.}, ``An open-source hw-sw co-development framework enabling efficient multi-accelerator systems,'' in \emph{IEEE/ACM International Symposium on Low Power Electronics and Design (ISLPED)}, 2025, pp. 1--7.

\bibitem{yi2025datamaestro}
X.~Yi \emph{et~al.}, ``Datamaestro: A versatile and efficient data streaming engine bringing decoupled memory access to dataflow accelerators,'' in \emph{62nd ACM/IEEE Design Automation Conference (DAC)}, 2025, pp. 1--7.

\bibitem{ST}
V.~Camus \emph{et~al.}, ``Review and benchmarking of precision-scalable multiply-accumulate unit architectures for embedded neural-network processing,'' \emph{IEEE Journal on Emerging and Selected Topics in Circuits and Systems (JETCAS)}, vol.~9, no.~4, pp. 697--711, 2019.

\bibitem{early-accum}
D.~R. Lutz \emph{et~al.}, ``Fused fp8 4-way dot product with scaling and fp32 accumulation,'' in \emph{2024 IEEE 31st Symposium on Computer Arithmetic (ARITH)}, 2024, pp. 40--47.

\bibitem{zaruba2020snitch}
F.~Zaruba \emph{et~al.}, ``Snitch: A tiny pseudo dual-issue processor for area and energy efficient execution of floating-point intensive workloads,'' \emph{IEEE Transactions on Computers (TC)}, vol.~70, no.~11, pp. 1845--1860, 2020.

\bibitem{Mant_accuracy}
\BIBentryALTinterwordspacing
Y.~Zhang \emph{et~al.}, ``Reduced precision checking to detect errors in floating point arithmetic,'' 2015. [Online]. Available: \url{https://arxiv.org/abs/1510.01145}
\BIBentrySTDinterwordspacing

\bibitem{moons201714}
B.~Moons \emph{et~al.}, ``14.5 envision: A 0.26-to-10tops/w subword-parallel dynamic-voltage-accuracy-frequency-scalable convolutional neural network processor in 28nm fdsoi,'' in \emph{2017 IEEE International Solid-State Circuits Conference (ISSCC)}.\hskip 1em plus 0.5em minus 0.4em\relax IEEE, 2017, pp. 246--247.

\bibitem{ueyoshi2022diana}
K.~Ueyoshi \emph{et~al.}, ``Diana: An end-to-end energy-efficient digital and analog hybrid neural network soc,'' in \emph{2022 IEEE International Solid-State Circuits Conference (ISSCC)}, vol.~65.\hskip 1em plus 0.5em minus 0.4em\relax IEEE, 2022, pp. 1--3.

\bibitem{dacapo}
Y.~Kim \emph{et~al.}, ``Dacapo: Accelerating continuous learning in autonomous systems for video analytics,'' in \emph{ACM/IEEE 51st Annual International Symposium on Computer Architecture (ISCA)}, 2024, pp. 1246--1261.

\bibitem{noune20228bitnumericalformatsdeep}
\BIBentryALTinterwordspacing
B.~Noune \emph{et~al.}, ``8-bit numerical formats for deep neural networks,'' 2022. [Online]. Available: \url{https://arxiv.org/abs/2206.02915}
\BIBentrySTDinterwordspacing

\bibitem{PS_MX_repo}
{KULeuven-MICAS}, ``Precision-scalable\_mx,'' \url{https://github.com/KULeuven-MICAS/Precision-Scalable_MX}, 2025, accessed: 2025-11-07.

\end{thebibliography}

\end{document}